\documentstyle[preprint,aps]{revtex}

\let\e=\epsilon
\let\k=\kappa\let\l=\lambda
\let\m=\mu\let\n=\nu\let\r=\rho
\let\s=\sigma\let\t=\tau
\let\w=\omega
\let\G=\Gamma
\let\D=\Delta\let\F=\Phi

\newcommand{\be}{\begin{equation}}
\newcommand{\ee}{\end{equation}}
\newcommand{\bea}{\begin{eqnarray}}
\newcommand{\eea}{\end{eqnarray}}

\newcommand{\del}{\partial}

\newcommand{\nbox}{{\,\lower0.9pt\vbox{\hrule \hbox{\vrule height 0.2 cm
\hskip
0.2 cm \vrule height 0.2 cm}\hrule}\,}}
\begin{document}
\tightenlines
\begin{titlepage}
\begin{center}
\vskip .2in
\hfill
\vbox{
    \halign{#\hfil         \cr
	   hep-th/9803082\cr
           SLAC-PUB-7767 \cr
           March 1998    \cr
           }  
      }   
\vskip 0.5cm
\begin{center}
{\large \bf Two-Form Fields and the Gauge Theory }\\
{\large \bf Description of Black Holes}\\
\end{center}
\vskip .2in
{\bf Arvind Rajaraman} \footnote{e-mail address:
arvindra@dormouse.stanford.edu}\footnote{Supported in part by the
Department of
Energy
under contract no. DE-AC03-76SF00515.}\\
\vskip .25in
{\em
Stanford Linear Accelerator Center,
     Stanford, CA 94309 }
\vskip 1cm
\end{center}
\begin{abstract}
We calculate the absorption cross section on a black threebrane of
two-form
perturbations polarized along the brane. The equations are coupled and
we decouple them for s-wave perturbations. The Hawking rate is
suppressed at low energies, and this is shown to
be reflected in the gauge theory by a coupling to a higher
dimension operator.
 \end{abstract}
\end{titlepage}
\newpage

\newpage
\section{Introduction}

It has been proposed recently that the large N limit of
maximally supersymmetric SU(N) Yang-Mills theory may
be described by supergravity on an AdS background \cite{Malda}. This
proposal was motivated by the agreement between
computations of Hawking radiation from black threebranes
and the corresponding expectation from gauge theory
\cite{kleb1,kleb2,kleb3,kleb4,Igor,newkleb}.

Based on this conjecture, Gubser, Klebanov and Polyakov
\cite{Igor} and Witten\cite{Witten} have given a concrete
proposal for how to relate correlation functions in the
gauge theory to supergravity computations. In their approach,
one calculates the supergravity action in the AdS space
subject to certain boundary conditions on the fields. The
boundary conditions are treated as the sources for
operators on the boundary. One can then read off the correlation
function of these operators from the supergravity action.

For practical calculations, one would like to know the solutions
at linearized level (at least) for all the bulk fields. This is the
first step to being able to compute multipoint correlation
functions in the gauge theory.

In this note, we shall  consider perturbations
of two-form
potentials which are polarized parallel to the brane. These fields
have coupled equations of motion (as was pointed out in
\cite{kleb3}.) We are able to decouple these equations in
the case of s-wave perturbations, and extract the absorption
cross section for quanta of these fields incident on the
black hole.

We find that the absorption cross-section for these fields is
suppressed at small frequencies relative to minimally coupled
scalars. This is somewhat surprising because these scalars
are not fixed in the sense of \cite{Kall,Kol} and therefore
are not expected to have suppressed absorption rates.

We find that the suppression of the absorption rate
is reflected in the gauge theory in that these scalars
are coupled to higher dimensional operators in the gauge theory,
which naturally leads to lower absorption rates.
One can formulate a conformally invariant coupling (along the
lines of \cite{Witten}) to describe this interaction. The
results are  in agreement with the semiclassical calculation.

Related issues have been discussed in
\cite{0,1,2,3,4,KS,6,65,Shiraz,Ofer,Jason1,Jason2,7,8,9,LNV,AV,CCDFFT,FFZ}.

\section{The semiclassical analysis}
\subsection{The black hole}
The black hole background is defined by the metric
\bea
ds^2&=&H^{-1/2}(-dt^2+dx_adx^a)+H^{1/2}(dx_i^2)\nonumber\\
H&=&1 + {R^4\over r^4}
\eea
where $a=1,2,3$ labels the coordinates parallel to the brane, $i=4\cdots
9$
labels the
coordinates perpendicular to the brane.

The four-form field strength is
\be
F_{0123r}=H^{-2}\left({R^4\over r^5}\right)
\ee

We are considering waves of two-form potentials. The relevant field
equations
at
the linearized level are \cite{Schwarz}
\bea
\nabla^\mu H_{\m\n\r}=\left({2\over
3}\right)F_{\n\r\k\t\s}F^{\k\t\s}\nonumber\\
\nabla^\mu F_{\m\n\r}=-\left({2\over 3}\right)F_{\n\r\k\t\s}H^{\k\t\s}
\eea
where we have denoted the NSNS two-form field strength as $H_{\m\n\r}$,
and the
RR two form field strength as $F_{\m\n\r}$. We shall denote the
corresponding
potentials as $B_{\m\n}$ and $A_{\m\n}$ respectively.

The above equations show that the perturbations of the two two-form
potentials
are
mixed. In particular, a perturbation of $A_{12}=\F$ mixes with
perturbations
of $H_{03r}$.  In the
case of s-waves, i.e. when there is no angular dependence, these equations
can be decoupled, and an equation for $\F$ can be obtained.
By symmetry, similar equations can be obtained for
$A_{13},A_{23},B_{12},B_{13}$ and $B_{23}.$

\subsection{Decoupling the equations of motion}
We start with the equation (we shall always assume the diagonal form of
the
metric)
\bea
{1\over
\sqrt{g}g^{rr}g^{33}}\del_0(\sqrt{g}g^{00}g^{33}g^{rr}H_{0r3})=\left({2\over
3}\right)F_{r3\m\n\r}F^{\m\n\r}\nonumber\\
=4F_{r3012}F^{012}
\eea
which gives (assuming everything goes as  $e^{-i\w t}$, we set
$\del_0=-i\w$)
\be
(i\w)g^{00}H_{0r3}=\left({4\over
3}\right)F_{r3012}g^{11}g^{22}g^{00}(i\w)\F
\ee
that is,
\be
H_{0r3}=\left({4\over 3}\right)F_{r3012}g^{11}g^{22}\F\label{eq:Fsoln}
\ee

We now turn to the equation
\be
{1\over \sqrt{g}g^{11}g^{22}}\del_r(\sqrt{g}g^{rr}g^{11}g^{22}\del_r\F)-
\w^2g^{00}\F=12 F_{120r3}H^{0r3}\nonumber
\ee
Using (\ref{eq:Fsoln}), we can simplify  this to
\bea
\del_r(Hr^5\del_r\F)+\w^2 H^{1/2}\F=16 {R^8\over r^5H}\F
\eea

\subsection{Solving the equations}

We will solve this in various regions.

 Far from the horizon $(r\gg R)$, we can set $H=1$, and $R=0$. The
equation is then
\bea
{1\over r^5}\del_r(r^5\del_r\F)+\w^2\F=0\nonumber
\eea
Using the standard substitution $\F=r^{-5/2}\G$,
we get
\bea
\left(\del_r^2-{15\over 4r^2}+\w^2\right)\G=0\nonumber
\eea
with the usual solution \cite{kleb2}
\bea
\F=c_1r^{-2}J_{2}(\w r)+c_2r^{-2}N_{2}(\w r)
\eea

In the intermediate region $(r\sim R \ll \w^{-1})$, we set $\w=0$. The
equation
is then
\bea
\del_r(Hr^5\del_r\F)=16 {R^8\over r^5H}\F\nonumber
\eea
with the solution
\bea
\F= c_3H+c_4H^{-1}
\eea

Finally, near the horizon $(r\ll R)$, we approximate
\bea
H={R^4\over r^4}\nonumber
\eea
The equation then becomes
\bea
r\del_r(r\del_r\F)+\left({\w^2R^4\over r^2}-16\right)\F=0\nonumber
\eea
with the solution
\be
\F=J_4\left( {\w R^2\over r}\right)+ i N_4\left( {\w R^2\over r}\right)
\ee
where we have chosen the solution for an ingoing wave.

To match the intermediate solution, we need
\bea
c_3={\w^4R^4\over 16\G(5)}\qquad
c_4={-96i\over \pi(\w R)^4}
\eea

Matching the intermediate to the outer solution, we get
\bea
c_1=\left({8\over \w^2}\right)(c_3+c_4)
\qquad
c_2=-\left({\w^2R^4\pi\over 4}\right)(c_3-c_4)
\eea

The solution for large r tends to $c_1 cos(\w r) +c_2sin(\w r)$.
The absorption cross section is then
\bea
A=1-\|{c_2+ic_1\over c_2-ic_1}\|^2\nonumber
\eea
which is easily evaluated to be
\bea
A=\left({\pi^2\over 2^{14}3^2}\right)(\w R)^{12}\nonumber
\eea
The cross-section is then obtained by the formula
\bea
\s={32\pi^2\over \w^5}A=\left({\pi^4\over
2^{9}3^2}\right)\w^7R^{12}\label{eq:absresult}
\eea

It may seem odd that we get a cross-section that goes to zero
more quickly than $\w^3$ (the behaviour exhibited by minimal scalars),
which is
reminiscent of the behaviour
of fixed scalars and intermediate scalars\cite{Kol,CGKT,Inter}. This may
be
unexpected since the scalar we are
considering is not expected to be fixed (in the sense of \cite{Kall}),
since it
can take on any value in the black hole background. It
nevertheless has a Hawking rate that is suppressed, essentially because
in the near horizon region it has an effective mass term similar to
the effective mass term of fixed scalars.
The presence of this  effective mass term is confirmed by
the analysis of \cite{Kim}, who have worked out the wave equations
of all the supergravity fields in an AdS background.

In the next section, we shall see that the suppression of the Hawking
rate is reflected in the gauge theory by the fact that the scalar we
are considering couples to a higher dimension operator on the
threebrane worldvolume.

\section{The gauge theory analysis}

We now turn to the extraction of gauge theory correlators from the
absorption amplitudes. This is a straightforward extension of the
analysis of  \cite{Igor,Witten} .We shall attempt to clarify the
relation between the two procedures.

We will focus on the near horizon equation
\bea
z\del_z(z\del_z\F)+\left(\w^2z^2-16\right)\F=0\label{eq:near}
\eea
where we have defined
\bea
z={ R^2\over r}\nonumber
\eea

We analytically continue to spacelike momenta, in which region the
solutions are $K_4(\w z)$ and $I_4(\w z)$ . We keep
the solution $K_4(\w z)$, which is the one which decays exponentially
at the horizon $z\rightarrow \infty$.

The idea of  \cite{Igor,Witten} is that one solves the above equation for
a
given choice
of boundary conditions for small $z$, which is taken to be the
value of the field on the boundary. We then treat the boundary field
as the source for an operator ${\cal O}$ which lives only on the
boundary. The Green's functions of ${\cal O}$ are then generated by the
functional obtained by substituting the full solution of (\ref{eq:near})
into
the
supergravity action.

In this case, the full solution is $K_4(\w z)$, which for small
$z$ diverges as $z^{-{4}}$. Accordingly, we need to specify a cutoff. The
problem is then that the boundary value is highly sensitive to
the choice of cutoff.

It is natural, therefore, to associate the boundary field not to $\F$
directly, but rather to the boundary value of $\F_0=z^4\F$. This is
stable in the sense that if we move the cutoff from $z=z_0$ to (say)
$z=2z_0$,
the
value of $\F_0$ does not change drastically. This is the same
setup as in section 2.5  of  \cite{Witten}.

We can then proceed as before, by coupling the field $\F_0$ to a
boundary operator ${\cal O}$, and extracting the correlation functions of
${\cal O}$ from the supergravity action.

Let us see explicitly how this works for the case of a  massive scalar.
We shall differ slightly from the method of \cite{Igor}, in that we
shall set $\k=1$, and explicitly follow all factors of $R$.

The
equation of motion is
\bea
z^5\del_z\left( {1\over
z^3}\del_z\chi\right)-z^2\w^2\chi-m^2\chi=0\nonumber
\eea
which has the solution

\bea
\chi= z^2K_\n(wz)\qquad \n^2=m^2+4\nonumber
\eea
In \cite{Igor}, the boundary condition chosen was
\bea
\chi\sim1 \qquad at \quad z=R\nonumber
\eea
We shall modify this condition in our case to
\bea
\chi\sim{R^{\n-2}\over z^{\n-2}} \qquad \nonumber
\eea
for small $\w z$,
which fixes the solution to be
\bea
\chi= R^{\n-2}\w^\n z^2K_\n(wz)\nonumber
\eea

We now substitute this solution into the supergravity action. As shown in
\cite{Igor,Witten}, this can be reduced to a surface term at the boundary
$z=R$,
\bea
I[\chi]\sim R^8\left[ \left({1\over z^3}\right) \chi\del_z
\chi\right]_R\label{eq:fluxeqn}
\eea
To extract the absorption cross-section, we need the nonanalytic part
of $I(\chi)$as in \cite{kleb4}, which is provided by the logarithm in the
expansion of $K_\n(z)$,
\bea
K_\n(z)\sim 2^{n-1}\G(n)z^{-\n}(1+\cdots)+(-)^{n+1}\left({1\over
2^n\G(n+1)}\right)ln\left({1\over 2}z\right)(z^\n+\cdots)\nonumber
\eea
where $\cdots$ represent higher orders in $z$. The leading nonanalytic
term in (\ref{eq:fluxeqn}) then scales as
\bea
R^8R^{2\n-4}\left({1\over
z^3}\right)(\w^{2\n}z^4)(\w^{-n}z^{-\n-1})ln(wz)(\w
z)^\n\sim R^{2\n+4}\w^{2\n}ln(wz)\nonumber
\eea
Upon Fourier transforming to position space, we find that the two point
function of the boundary operator scales as
\bea
\langle{\cal O}(x){\cal O}(0)\rangle\sim \del^{2\n}{1\over x^4}\nonumber
\eea
indicating that the dimension of the operator is
\bea
\D=2+\n=2+\sqrt{4+m^2}
\eea
in exact agreement with \cite{Witten}.

The reason that the coupling is still conformal is that the boundary
value $\chi_0$ has now acquired a dimension. In this case, this can be
seen in that if we shift the position of the boundary from $z=R$ to
$z=\l R$, the relation between $\chi$ and $\chi_0$ changes from
\bea
\chi_0= r^{\n-2}\chi\nonumber
\eea
to
\bea
\chi_0= \l^{\n-2}r^{\n-2}\chi\nonumber
\eea
Since $\chi$ itself was a canonical scalar field, $\chi_0$  is not (as
otherwise
the relation would not change under this rescaling.) In fact, it is a
conformal
density of dimension $2-\n$\cite{Witten}.
The coupling $\chi_0{\cal O}$ therefore has dimension 4, and is
a conformally invariant term. This is in spite of the fact that we
have introduced a higher dimension operator which will result
in a suppressed Hawking rate.

It is simple to repeat this analysis for massive p-forms. The equation of
motion is
\bea
z^{5-2p}\del_z\left( {1\over
z^{3-2p}}\del_z\chi^{(p)}\right)+z^2\w^2\chi^{(p)}-m^2\chi^{(p)}=0\nonumber
\eea
with the solution
\bea
\chi^{(p)}=R^{\n+p-2} \w^\n z^{2-p}K_\n(wz)\qquad
\n^2=m^2+(2-p)^2\label{eq:soln2}
\eea
where we have normalized the solution to go as
\bea
\chi^{(p)}\sim {R^{\n+p-2}\over z^{\n+p-2}} \nonumber
\eea
for small $\w z$.

We can reduce the action to a surface term as before
\bea
I[\chi^{(p)}]\propto\left[ \left({z^{2p-3}\over R^{2p-8}}\right)
\chi^{(p)}\del_z \chi^{(p)}\right]_R\label{eq:fluxeqn2}
\eea

Approximating the behaviour of $\chi^{(p)}$ at small $z$ as before, we
find the leading nonanalytic term
\bea
I\sim {z^{2p-3}\over
R^{2p-8}}(R^{2\n+2p-4}\w^{2\n}z^{4-2p})(\w^{-\n}z^{-\n-1})(\w^\n z^\n
ln(wz))\sim R^{2\n-4}\w^{2\n}ln(wR)\nonumber
\eea

Hence the dimension of the operator on the boundary is
\bea
\D=2+\n=2+\sqrt{m^2+(2-p)^2}
\eea
which is the correct result \cite{Witten,Ofer}, as the $m^2$ refers not to
the
eigenvalue of the
Laplacian, but to the eigenvalue of the Maxwell operator ($\tilde{m}^2$ in
\cite{Ofer}.)

There is however
a puzzle in the comparison of the gauge theory to the semiclassical
calculation.
The problem is that
in (\ref{eq:fluxeqn2}), if we explicitly substitute the solution
(\ref{eq:soln2}), the leading
nonanalytic term cancels! This is because
\bea
I\propto \del_z\chi^2  \nonumber
\eea
and in $\chi^2$, the leading coefficient of $ln(\w z)$ is $z^0$. Hence,
upon
differentiation,
we find that the nonanalytic term $\ln(\w)$ disappears.
Another way of saying this is that if we treat $p$ as a continuous
variable,
the
coefficient of the action is proportional to $(p-2)$ and hence vanishes
for
two-forms.

It is
clear that the true answer in the gauge theory cannot be zero, since the
absorption cross section is nonzero. Also, the
procedure of \cite{Witten} does not seem to give zero for this case.
This may be a problem of our normalization. If one treats $p$ as
continuous, it is possible that the normalization of $\chi^{(p)}$ should
be
taken to involve inverse powers of $(p-2)$ which will cancel
the apparent zero in the above expression. Other possibilities may exist.
We will treat this as an overall coefficient in the correlation function
that we cannot determine, since we cannot normalize the operators
unambiguously.

In particular, in the case we are considering, we have $p=2$, and
$m^2=16$,
which
yields $\D=6$. Hence we have a coupling to a dimension 6 operator.
The exact form of this operator has been discussed in more detail in
\cite{FFZ}.

We also find that
\bea
I\sim R^{12}\w^8ln(wR)\nonumber
\eea
and since the cross section is related to the discontinuity of the above
function
near $\w=0$, we find (from \cite{kleb4})
\bea
\s\sim {i\over \w}R^{12}\w^8( ln(-s+i\e)-ln(s-i\e) ) \sim \w^7
R^{12}\nonumber
\eea
which agrees with (\ref{eq:absresult}).

We therefore  get results in agreement with the semiclassical
calculation. The
exact coefficient is, however, undetermined.
We  emphasize that this is because the exact normalization of the
operators
has not been fixed.
It may be necessary to calculate a three-point correlation function in
order
to resolve the ambiguity.

In conclusion, we have extracted the absorption rate for a two-form field
incident on a black threebrane. We have shown that the Hawking rate is
proportional to
$\w^7$, a fact which follows from a coupling to a dimension 6 operator
on the brane world volume. We have thus found that a non-fixed scalar can
also
have a
suppressed cross-section.

\section{Acknowledgements}

We would like to thank I. Klebanov and J. Rahmfeld  for  discussions.

This work was supported in part by
the Department of Energy
under contract no. DE-AC03-76SF00515.

\end{document}